\documentclass[3p,times,twocolumn]{elsarticle}
\usepackage{amssymb}
\usepackage[figuresright]{rotating}

\newcommand{\bc}{\begin{center}}
\newcommand{\ec}{\end{center}}
\newcommand{\be}{\begin{equation}}
\newcommand{\ee}{\end{equation}}
\newcommand{\bea}{\begin{eqnarray}}
\newcommand{\eea}{\end{eqnarray}}
\newcommand{\ba}{\begin{array}}
\newcommand{\ea}{\end{array}}
\newcommand{\lb}{\label}
\newcommand{\rf}{\ref}
\newcommand{\bfg}{\begin{figure}[htbp]}
\newcommand{\efg}{\end{figure}}

\begin{document}

\begin{frontmatter}




\title{Two-point gauge invariant quark Green's functions \protect \\
with polygonal phase factor lines}


\author{H. Sazdjian}

\address{Institut de Physique Nucl\'eaire, CNRS/IN2P3,\\
Universit\'e Paris-Sud, F-91405 Orsay, France}

\ead{sazdjian@ipno.in2p3.fr}

\begin{abstract}
Polygonal lines are used for the paths of the gluon field phase
factors entering in the definition of gauge invariant quark Green's 
functions. This allows classification of the Green's functions according 
to the number of segments the polygonal lines contain. Functional relations
are established between Green's functions with polygonal lines with
different numbers of segments. An integro\-differen\-tial equation is 
obtained for the quark two-point Green's function with a path along a
single straight line segment where the kernels are represented by a series 
of Wilson loop averages along polygonal contours. The equation is exactly 
and analytically solved in the case of two-dimensional QCD in the large-$N_c$ 
limit. The solution displays generation of an infinite number of dynamical 
quark masses accompanied with branch point singularities that are stronger 
than simple poles. An approximation scheme, based on the counting of 
functional derivatives of Wilson loops, is proposed for the resolution of 
the equation in four dimensions.  
\end{abstract}


\end{frontmatter}

% \linenumbers

\section{Introduction} \lb{s1}

Path-dependent phase factors are the natural ingredients in gauge theories
for the description of parallel transport of gauge covariant quantities
in integrated form \cite{Mandelstam:1968hz,Nambu:1978bd}. They also allow 
for the construction of gauge invariant Green's functions,
which are the adequate tools for the study of the physical 
properties of the theory. The Wilson loop \cite{Wilson:1974sk}, which 
corresponds to a phase factor along a closed contour, is used to set
up a criterion for the recognition of the confinement of quarks in QCD 
\cite{Wilson:1974sk,Brown:1979ya,Kogut:1982ds}. Properties of Wilson 
loops were thoroughly studied in the past \cite{Polyakov:1980ca,
Makeenko:1979pb,Makeenko:1980wr,Makeenko:1980vm,Dotsenko:1979wb,Brandt:1981kf} 
and applications to bound states of quarks were considered 
\cite{Eichten:1980mw,Barchielli:1986zs,Barchielli:1988zp,Simonov:1987rn,
Brambilla:1993zw,Dubin:1994vn,Brambilla:1997ky,Brambilla:2000gk,Pineda:2000sz, 
Jugeau:2003df}.  
\par
On the other hand, approaches using gauge invariant correlators meet
difficulties due to the extended nature of the phase factors and
could not up to now provide a complete systematic procedure of solving
the theory. In spite of these difficulties, the advantages one might
expect from a gauge invariant approach merit continuation of the efforts
that are undertaken. Gauge invariant quantities are expected to have 
an infrared safe behavior, free of artificial or unphysical singularities.
For this reason, they are better suited to explore the nonpertubative
regime of the theory; in QCD, this mainly concerns the occurrence of
confinement. Also, the knowledge of gauge invariant wave functions of
bound states facilitates calculations of matrix elements of operators
involving phase factors.
\par
The present talk is a summary of recent work of the author 
\cite{Sazdjian:2007ng,Sazdjian:2010ku} trying to deduce exact
integro\-differen\-tial equations for two-point quark gauge invariant 
Green's functions (2PQGIGF), in analogy with the Dyson-Schwinger equations 
of ordinary Green's functions 
\cite{Dyson:1949ha,Schwinger:1951ex,Schwinger:1951hq,
Alkofer:2000wg,Fischer:2006ub}. The method of approach is based on
the use of polygonal lines for the paths of the phase factors.
Polygonal lines are of particular interest since they can be
decomposed as a succession of straight line segments with mutual 
junction points. Straight line segments are Lorentz invariant in form
and have an unambiguous limit when the two end points approach each 
other. Polygonal lines can be classified according to the number of
segments or sides they contain, which in turn is reflected on the
2PQGIGFs.  
\par

\section{Green's functions with polygonal lines} \lb{s2}  

Let $U(y,x)$ be a path-ordered phase factor along an oriented 
straight line segment going from $x$ to $y$. A displacement of one
end point of the rigid segment, while the other end point is
fixed, generates also a displacement of the interior points of the
segment. This defines a rigid path displacement. Parametrizing 
the interior points of the segment with a linear parameter $\lambda$
varying between $0$ and $1$, such that $z(\lambda)=\lambda y+(1-\lambda)x$,
the rigid path derivative operations with respect to $y$ or $x$ yield
\cite{Mandelstam:1968hz,Nambu:1978bd,Corrigan:1978zg,Durand:1979sw}
\bea \lb{2e1}
\lefteqn{\hspace{-0.25 cm} \frac{\partial U(y,x)}{\partial y^{\alpha}}
=-igA_{\alpha}(y)U(y,x)+ig(y-x)^{\beta}}\nonumber \\
& &{\hspace{-0.25cm} \times\int_0^1d\lambda\lambda U(y,z(\lambda))
F_{\beta\alpha}(z(\lambda))U(z(\lambda),x),}
\eea
\bea \lb{2e2}
\lefteqn{\hspace{-0.25 cm}\frac{\partial U(y,x)}{\partial x^{\alpha}}
=+igU(y,x)A_{\alpha}(x)+ig(y-x)^{\beta}}\nonumber \\
& &{\hspace{-.25cm} \times\int_0^1d\lambda(1-\lambda) U(y,z(\lambda))
F_{\beta\alpha}(z(\lambda))U(z(\lambda),x),}\nonumber \\
& &
\eea
where $A$ is the gluon potential, $F$ its field strength and $g$ 
the coupling constant.
\par
In gauge invariant quantities, the end point contributions of the
segments are usually cancelled by other neighboring point
contributions and one remains only with the interior point
contributions of the segments, represented by the integrals above.
We introduce for them a shorthand notation:
\bea 
\lb{2e3}  
\lefteqn{\frac{\bar\delta U(y,x)}{\bar\delta y^{\alpha +}}=
ig(y-x)^{\beta}\int_0^1d\lambda\lambda U(y,z(\lambda))}
\nonumber \\
& &\ \ \ \ \ \ \ \ \ \ \times F_{\beta\alpha}(z(\lambda))U(z(\lambda),x),
\eea
\bea
\lb{2e4}
\lefteqn{\frac{\bar\delta U(y,x)}{\bar\delta x^{\alpha -}}=
ig(y-x)^{\beta}\int_0^1d\lambda(1-\lambda) U(y,z(\lambda))}
\nonumber \\
& &\ \ \ \ \ \ \ \ \ \ \times F_{\beta\alpha}(z(\lambda))U(z(\lambda),x).
\eea
The superscript $+$ or $-$ on the derivative variable takes account 
of the orientation on the segment and specifies, in the case of
joined segments, the segment on which the derivative acts.
\par
The vacuum expectation value (or vacuum average) of a Wilson loop 
along a contour $C$ will be designated by $W(C)$. In the case of 
a polygonal contour $C_n$, with $n$ segments and $n$ junction
points $x_1$, $x_2$, $\ldots$, $x_n$, it will be designated by
$W_n$ and represented as an exponential functional 
\cite{Makeenko:1980wr,Dotsenko:1979wb}:
\be \lb{2e5}
\hspace{-.5 cm}
W_n=W(x_n,x_{n-1},\ldots,x_1)=e^{F_n(x_n,x_{n-1},\ldots,x_1)}
=e^{F_n}.
\ee
\par
The 2PGIQGF with a phase factor along a polygonal line composed of 
$n$ segments and $(n-1)$ junction points is designated by $S_{(n)}$:
\bea \lb{2e6}
\lefteqn{S_{(n)}(x,x';t_{n-1},\ldots,t_1)=-\frac{1}{N_c}\langle 
\overline\psi(x')U(x',t_{n-1})}\nonumber \\
& &\ \ \ \ \ \ \ \times U(t_{n-1},t_{n-2})\ldots U(t_1,x)\psi(x)\rangle,
\eea
the quark fields, with mass parameter $m$, belonging to the 
fundamental representation of the color gauge group $SU(N_c)$ and 
the vacuum expectation value being defined in the path integral 
formalism. (Spinor indices are omitted and the color indices are 
implicitly summed.) The simplest such function is $S_{(1)}$, having 
a phase factor along a straight line segment:
\be \lb{2e7} 
\hspace{-0.5cm}
S_{(1)}(x,x')\equiv S(x,x')=-\frac{1}{N_c}\langle 
\overline\psi(x')U(x',x)\psi(x)\rangle.
\ee
(We shall generally omit the index 1 from that function.)
\par
For the internal parts of rigid path derivatives, we have 
definitions of the type
\bea \lb{2e8}
\lefteqn{\frac{\bar\delta S_{(n)}(x,x';t_{n-1},\ldots,t_1)}
{\bar\delta x^{\mu -}}=-\frac{1}{N_c}\langle 
\overline\psi(x')U(x',t_{n-1})}\nonumber \\
& &\ \ \ \ \ \ \ \ \times U(t_{n-1},t_{n-2})\ldots 
\frac{\bar\delta U(t_1,x)}{\bar\delta x^{\mu -}}\psi(x)\rangle.
\eea
\par

\section{Integrodifferential equation} \lb{s3}

The above Green's functions satisfy the following equations
of motion concerning the quark field variables:
\bea \lb{3e1}
\lefteqn{\hspace{-.5 cm} (i\gamma.\partial_{(x)}-m)S_{(n)}(x,x';t_{n-1},
\ldots,t_1)=i\delta^4(x-x')}\nonumber \\
& &{\hspace{-0.5cm} \times e^{F_n(x,t_{n-1},\ldots,t_1)}
+i\gamma^{\mu}\frac{\bar\delta S_{(n)}(x,x';t_{n-1},
\ldots,t_1)}{\bar\delta x^{\mu -}}},
\eea
which become for $n=1$
\be \lb{3e2}
\hspace{-0.65cm}
(i\gamma.\partial_{(x)}-m)\ S(x,x')=i\delta^4(x-x')+
i\gamma^{\mu}\frac{\bar\delta S(x,x')}{\bar\delta x^{\mu -}}.
\ee
\par
Multiplying Eq. (\rf{3e1}) with $S(t_1,x)$ and integrating with respect to
$x$, one obtains functional relations between various 2PGIQGFs. A
typical such relation is:
\bea \lb{3e3}
\lefteqn{S_{(n)}(x,x';t_{n-1},\ldots,t_1)=S(x,x')\ 
e^{F_{n+1}(x',t_{n-1},\ldots,t_1,x)}}\nonumber \\
& &\ \ \ \ \ +\Big(\frac{\bar\delta S(x,y_1)}{\bar\delta y_1^{\alpha_1 +}}
+S(x,y_1)\frac{\bar\delta}{\bar\delta y_1^{\alpha_1 -}}\Big)\nonumber \\ 
& &\ \ \ \ \ \ \ \ \ \times S_{(n+1)}(y_1,x';t_{n-1},\ldots,t_1,x).
\eea
[Integrations on intermediate variables are implicit and generally 
are not written throughout this paper. Here, $y_1$ is an 
integration variable.] 
\par
Equation (\rf{3e3}) expresses $S_{(n)}$ in terms of two quantities:
the first one, which plays the role of a driving term, contains the
simplest 2PGIQGF, with one straight line segment, together with a 
Wilson loop average along a polygonal contour with $(n+1)$ segments;
the second term, which appears as a corrective term, is represented 
by the contribution of a higher-index 2PGIQGF. However, since this 
equation is valid for any $n\ge 1$, one can use it again in its 
right-hand side for $S_{(n+1)}$. One therefore generates an iterative 
procedure that eliminates successively the higher-index 2PGIQGFs in 
terms of the lowest-index one, $S_{(1)}$. Assuming that the terms 
rejected to infinity are negligible, one ends up with a series where 
only $S_{(1)}$ appears together with Wilson loop averages along 
polygonal contours with an increasing number of sides and rigid path 
derivatives along the segments. This result shows that among the set 
of the 2PGIQGFs $S_{(n)}$, $n=1,2,\ldots$, it is only $S_{(1)}$, 
having a phase factor along one straight line segment, that is a 
genuine dynamical independent quantity. Higher-index 2PGIQGFs could 
in principle be eliminated in terms of $S_{(1)}$, together with 
polygonal Wilson loops and their rigid path derivatives.
\par
The construction of $S$ proceeds from the resolution of the equation 
of motion (\rf{3e2}). It is then necessary to evaluate the action
of the rigid path derivative on $S$ as it appears in the right-hand 
side of the equation. This is done by using again the functional
relations (\rf{3e3}), where the driving term of the right-hand side gives
the main contribution. Thus, the rigid path derivative acting along the 
segment $xt_1$ of $S_{(n)}$, acts in the right-hand side in the first
place on the logarithm of the Wilson loop average $F_{n+1}$; it also
acts on the remainder containing $S_{(n+1)}$. Using back Eq. (\rf{3e3}),
one obtains an equation for $\frac{\bar\delta S_{(n)}}{\bar\delta x^-}$
which expresses the latter as a product of 
$\frac{\bar\delta F_{n+1}}{\bar\delta x^-}$ with $S_{(n)}$ plus a
remainder containing the derivative of $S_{(n+1)}$. Continuing the
procedure, one factorizes in front of every $S_{(n')}$ ($n'>n$)
derivatives of Wilson loop averages. 
\par
Selecting in the above set of equations the case $n=1$, 
the equation of motion (\rf{3e2}) takes at the end the following form:
\bea \lb{3e4}
\lefteqn{\hspace{-0.5 cm}(i\gamma.\partial_{(x)}-m)\,S(x,x')
=i\delta^4(x-x')+i\gamma^{\mu}\Big\{K_{1\mu -}(x',x)}\nonumber \\
& &\times S(x,x')
+K_{2\mu -}(x',x,y_1)\,S_{(2)}(y_1,x';x)\nonumber \\
& &+\sum_{n=3}^{\infty}K_{n\mu -}(x',x,y_1,\ldots,y_{n-1})
\nonumber \\
& &\ \ \ \ \ \ \ \ \times S_{(n)}(y_{n-1},x';x,y_1,\ldots,y_{n-2})\Big\},
\eea
where the kernels $K_n$ ($n=1,2,\ldots$) contain Wilson loop averages
along polygonal contours, at most $(n+1)$-sided, and ($n-1$) 
2PGIQGFs $S$ and their derivative. The total number of derivatives contained 
in $K_n$ is $n$, each derivative acting on a different segment.
Once the Wilson loop averages and the various derivatives have been
evaluated and the high-index $S_{(n)}$s have been expressed in terms
of $S$, Eq. (\rf{3e4}) becomes an integro\-differen\-tial equation 
in $S$, which is the primary unknown quantity to be solved.
One observes in the right-hand side of the equation
the appearance of the whole set of 2PGIQGFs. Gluon 
propagators are replaced here by Wilson loop averages along polygonal 
contours and rigid path derivatives acting on them. This ensures gauge 
invariance of every term of the expansion. 
\par
One major difference of the integrals present in the right-hand side
of Eq. (\rf{3e4}) with those of the Dyson--Schwinger equation is the 
property that they are not of the convolution type. This is due to
the presence of the Wilson loops, whose contours pass by all points
of the accompanying terms and do not allow for a convolutive
factorization in $x$-space.
\par      
Equation (\rf{3e4}) can also be analyzed, at least 
superficially, from the viewpoint of a perturbative expansion. 
According to Eqs. (\rf{2e3}) and (\rf{2e4}), each derivative operator 
results in an insertion of the gluon field strength, leading to the 
appearance of a valence gluon, accompanied multiplicatively by the 
coupling constant. In the short-distance regime, where perturbative 
QCD should be applicable, a naive counting of the number of 
derivatives would give us an indication about the size of the 
corresponding term, the leading terms corresponding to those having 
the least number of derivatives. Here, perturbation theory would be
effected in the presence of the polygonal Wilson loops for each
term. At large distances, it is expected that Wilson loop averages
are saturated by minimal surfaces \cite{Makeenko:1980wr,Jugeau:2003df}.
Here also, increasing the number of derivatives would lead to
less dominant terms. It thus seems reasonable to assume, as a
starting hypothesis, that Eq. (\rf{3e4}) represents, on practical
grounds, a perturbative expansion. The first term of the
series, corresponding to a single derivative term, is null for
symmetry reasons. Therefore, the leading term of the series would be
represented by the two-derivative term with a Wilson loop average
along a triangular contour.
\par 
In that case, the interaction part of Eq. (\rf{3e2}) reduces to
the expression
\bea \lb{3e5} 
\lefteqn{\frac{\bar\delta S(x,x')}{\bar\delta x^{\mu -}}\simeq
-\int d^4y_1\,\frac{\bar\delta^2 F_3(x',x,y_1)}{\bar\delta x^{\mu -}
\bar\delta y_1^{\alpha_1 +}}}\nonumber \\
& &\ \ \ \ \ \times e^{F_3(x',x,y_1)}\,S(x,y_1)\,\gamma^{\alpha_1}\,
S(y_1,x'),
\eea
which provides the driving term of the kernel of the equation.
\par

\section{Two-dimensional QCD} \lb{s4}

The equations obtained in the previous sections remain also valid 
in two dimensions and could be analyzed more easily in that case.
Two-dimensional QCD in the large $N_c$ limit \cite{'tHooft:1973jz,
'tHooft:1974hx} provides a simplified framework for the study of 
the confinement properties which are expected to prevail also in 
four dimensions. Wilson loop averages can be explicitly calculated
\cite{Kazakov:1980zi,Kazakov:1980zj,Bralic:1980ra}: for simple
contours they are equal to the exponential of the areas enclosed
by the contours. In that case, the second-order derivative of the
logarithm of the Wilson loop average reduces to a two-dimensional
delta-function. Higher-order derivatives give zero, since they act 
on different segments of the polygonal contour. The case of 
overlapping self-intersecting surfaces, which give more complicated
expressions, should be analyzed separately. A detailed analysis 
suggests that the residual terms they produce are probably of
zero weight under the integrations that are involved. We assume
that hypothesis.    
\par
In the series of terms of Eq. (\rf{3e4}) it is only the second-order
derivative that survives [cf. Eq. (\rf{3e5})] and the 
integro\-differen\-tial equation takes the following (exact) expression 
\cite{Sazdjian:2010ku}:
\bea \lb{4e1}
\lefteqn{\hspace{-0.5cm} (i\gamma.\partial-m)S(x)=i\delta^2(x)
-\sigma\gamma^{\mu}(g_{\mu\alpha}g_{\nu\beta}-g_{\mu\beta}g_{\nu\alpha}) 
x^{\nu}x^{\beta}}\nonumber \\
& &\ \times\Big[\,\int_0^1d\lambda\,\lambda^2\,S((1-\lambda)x)
\gamma^{\alpha}S(\lambda x)\nonumber \\
& &\ \ \ \ \ +\int_1^{\infty}d\xi\,S((1-\xi)x)\gamma^{\alpha}S(\xi x)\,
\Big], 
\eea
where $\sigma$ is the string tension.
\par
The above equation can be analyzed by first passing to momentum
space. Designating by $S(p)$ the Fourier transform of $S(x)$, one
can decompose it into Lorentz invariant components:
\be \lb{4e2}
S(p)=\gamma.pF_1(p^2)+F_0(p^2).
\ee
The solution of Eq. (\rf{4e1}) can be searched for by using the 
analyticity properties of the 2PGIQGF, assuming that the quark and 
gluon fields satisfy the usual spectral and causality properties of 
quantum field theory \cite{Sazdjian:2007ng,Sazdjian:2010ku,
Wightman:1956zz,Schweber:1961zz,'tHooft:1973pz}.
It turns out that the equation can be solved exactly and in analytic 
form. The functions $F_1$ and $F_0$ are found having an infinite 
number of branch points located on the positive real axis of $p^2$
(timelike region), starting at thresholds $M_1^2$, $M_2^2$, $\ldots$,
$M_n^2$, $\ldots$, with fractional power singularities equal to
$-3/2$. Their expressions are \cite{Sazdjian:2010ku}, for complex $p^2$,
\bea 
\lb{4e3}
& &F_1(p^2)=-i\frac{\pi}{2\sigma}\sum_{n=1}^{\infty}\,b_n
\frac{1}{(M_n^2-p^2)^{3/2}},\\
\lb{4e4}
& &F_0(p^2)=i\frac{\pi}{2\sigma}\sum_{n=1}^{\infty}(-1)^nb_n
\frac{M_n}{(M_n^2-p^2)^{3/2}}.
\eea
The Green's function $S$ [Eq. (\rf{4e2})] then takes the form
\be \lb{4e5}
S(p)=-i\frac{\pi}{2\sigma}\sum_{n=1}^{\infty}\,b_n
\frac{(\gamma.p+(-1)^{n+1}M_n)}{(M_n^2-p^2)^{3/2}}.
\ee
The masses $M_n$ ($n=1,2,\ldots$) are positive, greater than the free 
quark mass $m$ and ordered according to increasing values. For massless 
quarks they remain positive. The masses $M_n$ and the coefficients $b_n$,
the latter being also positive, satisfy, for general $m$, an infinite 
set of algebraic equations that are solved numerically. Their asymptotic 
values, for large values of $n$ such that $n\gg m^2/(\pi\sigma)$, are       
\be \lb{4e6}
M_n^2\simeq \pi n\sigma,\ \ \ \ \ \ \ \ 
b_n\simeq \frac{\sigma^2}{M_n+(-1)^nm}.
\ee
The functions $(M_n^2-p^2)^{-3/2}$ are defined with cuts starting from
their branch points and going to $+\infty$ on the real axis; they are
real below their branch points on the real axis down to $-\infty$.
\par
The expressions $(\rf{4e3})$ and $(\rf{4e4})$ are represented by weakly
converging series. The high-energy behavior of the functions $F_1$
and $F_0$ is obtained with a detailed study of the asymptotic tails
of the series and the use of the asymptotic behaviors of the parameters
$M_n$ and $b_n$ [Eqs. (\rf{4e6})]. One finds that they behave as in
free field theories, which is here a trivial manifestation of 
asymptotic freedom \cite{Politzer:1976tv}:
\be \lb{4e7}
F_1(p^2)_{\stackrel{{\displaystyle =}}{p^2\rightarrow -\infty}}
\frac{i}{p^2},
\ee
\be \lb{4e8}
F_0(p^2)_{\stackrel{{\displaystyle =}}{p^2\rightarrow -\infty}}
\frac{im}{p^2},\ \ \ \mathrm{for}\ \ m\ne 0,
\ee
\be \lb{4e9}
F_0(p^2)_{\stackrel{{\displaystyle =}}{p^2\rightarrow -\infty}}
-\frac{4i\sigma F_0(x=0)}{(p^2)^2},\ \ \ \mathrm{for}\ \ m=0.
\ee
\par
In summary, the solution of Eq. (\rf{4e1}) is nonperturbative
and infrared finite. The masses $M_n$ are dynamically generated,
since they do not exist in the Lagrangian of the theory. They could
be interpreted as dynamical masses of quarks with, however, the
following particular features. First, they are infinite in number. 
Second, they do not appear as poles in the Green's function, but 
rather with stronger singularities. In $x$-space the latter do not 
produce finite plane waves at large distances and therefore quarks 
could not be observed as free asymptotic states. Nevertheless, the 
above singularities being gauge invariant should have physical 
significance and would show up in the infrared regions of physical 
processes involving quarks. Finally, the fact that they appear only 
in the timelike region of real $p^2$ is an indication that 
even in the nonperturbative regime the spectral and causality properties 
of quantum field theory are still satisfied 
\cite{Wightman:1956zz,Schweber:1961zz,'tHooft:1973pz}.
\par    
Expression (\rf{4e5}) of the Green's function $S$ can be 
interpreted as fitting a generalized form of the K\"all\'en-Lehmann
representation \cite{Sazdjian:2007ng,Sazdjian:2010ku,Kallen:1952zz,
Lehmann:1954xi}, where the denominator of the dispersive
integral has now a fractional power, while the spectral 
functions are saturated by an infinite series of dynamically 
generated single quark states with alternating parities. The latter
still satisfy Lehmann's positivity conditions \cite{Lehmann:1954xi}.
\par 

\section{Conclusion} \lb{s5}

The use of polygonal lines for the paths of the phase factors allow
a classification of the two-point quark Green's functions and a
systematic investigation of the properties of the latter through the 
functional relations they satisfy. An equation similar to the
Dyson-Schwinger equation has been obtained for the quark Green's 
function with a path made of one straight line segment, in which the 
kernels are represented by a series involving Wilson loop 
averages along closed polygonal contours with increasing complexity.
Arguments have been developed justifying the treatment of the series
perturbatively with respect to the number of functional derivatives
acting on the Wilson loops. In that case the leading term of the 
kernels is provided by the Wilson loop with the simplest contour,
corresponding to a triangle, with two functional derivatives.
\par
The above equation has been solved exactly and analytically in the
case of two-dimensional QCD in the large-$N_c$ limit. The solution
displays the presence of an infinite number of dynamically generated
quark masses, accompanied with branch point singularities of degree 
$-3/2$ (stronger than simple poles) located on the positive real axis 
of the momentum squared variable (timelike region). The qualitative 
feature that one deduces from these results is that in spite of the 
strong singularities that have emerged in the exact solution, quark 
and gluon fields continue satisfying the spectral and causality 
properties of quantum field theory.
\par
The resolution of the integrogrodifferential equation in the 
two-dimensional case provides a positive signal for the continuation 
of the analysis in four dimensions following similar lines.
\par 

\section*{Acknowledgements}

I thank Professor Daya S. Kulshreshtha and the Organizing Committee of 
the Light Cone Delhi 2012 Conference for the pleasant and stimulating 
atmosphere created during the Conference and for their warm hospitality.
This work is partially supported by the EU I3HP Project ''Study of
Strongly Interacting Matter'' (acronym HadronPhysics3, Grant
Agreement No. 283286).
\par  



%% References

\nocite{*}
\bibliographystyle{elsarticle-num}
\bibliography{giqgfppfl}

\begin{thebibliography}{10}
\expandafter\ifx\csname url\endcsname\relax
  \def\url#1{\texttt{#1}}\fi
\expandafter\ifx\csname urlprefix\endcsname\relax\def\urlprefix{URL }\fi
\expandafter\ifx\csname href\endcsname\relax
  \def\href#1#2{#2} \def\path#1{#1}\fi

\bibitem{Mandelstam:1968hz}
S.~Mandelstam, {Feynman rules for electromagnetic and Yang-Mills fields from
  the gauge independent field theoretic formalism}, Phys.Rev. 175 (1968) 1580.

\bibitem{Nambu:1978bd}
Y.~Nambu, {QCD and the string model}, Phys.Lett. B80 (1979) 372.

\bibitem{Wilson:1974sk}
K.~G. Wilson, {Confinement of quarks}, Phys.Rev. D10 (1974) 2445.

\bibitem{Brown:1979ya}
L.~S. Brown, W.~I. Weisberger, {Remarks on the static potential in quantum
  chromodynamics}, Phys.Rev. D20 (1979) 3239.

\bibitem{Kogut:1982ds}
J.~B. Kogut, {A review of the lattice gauge theory approach to quantum
  chromodynamics}, Rev.Mod.Phys. 55 (1983) 775.

\bibitem{Polyakov:1980ca}
A.~M. Polyakov, {Gauge fields as rings of glue}, Nucl.Phys. B164 (1980) 171.

\bibitem{Makeenko:1979pb}
Y.~Makeenko, A.~A. Migdal, {Exact equation for the loop average in multicolor
  QCD}, Phys.Lett. B88 (1979) 135.

\bibitem{Makeenko:1980wr}
Y.~Makeenko, A.~A. Migdal, {Self-consistent area law in QCD}, Phys.Lett. B97
  (1980) 253.

\bibitem{Makeenko:1980vm}
Y.~Makeenko, A.~A. Migdal, {Quantum chromodynamics as dynamics of loops},
  Nucl.Phys. B188 (1981) 269.

\bibitem{Dotsenko:1979wb}
V.~Dotsenko, S.~Vergeles, {Renormalizability of phase factors in non-Abelian
  gauge theory}, Nucl.Phys. B169 (1980) 527.

\bibitem{Brandt:1981kf}
R.~A. Brandt, F.~Neri, M.-a. Sato, {Renormalization of loop functions for all
  loops}, Phys.Rev. D24 (1981) 879.

\bibitem{Eichten:1980mw}
E.~Eichten, F.~Feinberg, {Spin dependent forces in QCD}, Phys.Rev. D23 (1981)
  2724.

\bibitem{Barchielli:1986zs}
A.~Barchielli, E.~Montaldi, G.~Prosperi, {On a systematic derivation of the
  quark-antiquark potential}, Nucl.Phys. B296 (1988) 625.

\bibitem{Barchielli:1988zp}
A.~Barchielli, N.~Brambilla, G.~Prosperi, {Relativistic corrections to the
  quark-antiquark potential and the quarkonium spectrum}, Nuovo Cim. A103
  (1990) 59.

\bibitem{Simonov:1987rn}
Y.~Simonov, {Vacuum background fields in QCD as a source of confinement},
  Nucl.Phys. B307 (1988) 512.

\bibitem{Brambilla:1993zw}
N.~Brambilla, P.~Consoli, G.~Prosperi, {A consistent derivation of the
  quark-antiquark and three quark potentials in a Wilson loop context},
  Phys.Rev. D50 (1994) 5878.
\newblock \href {http://arxiv.org/abs/hep-th/9401051}
  {\path{arXiv:hep-th/9401051}}.

\bibitem{Dubin:1994vn}
A.~Y. Dubin, A.~Kaidalov, Y.~Simonov, {Dynamical regimes of the QCD string with
  quarks}, Phys.Lett. B323 (1994) 41.

\bibitem{Brambilla:1997ky}
N.~Brambilla, A.~Vairo, {From the Feynman-Schwinger representation to the
  nonperturbative relativistic bound state interaction}, Phys.Rev. D56 (1997)
  1445.
\newblock \href {http://arxiv.org/abs/hep-ph/9703378}
  {\path{arXiv:hep-ph/9703378}}.

\bibitem{Brambilla:2000gk}
N.~Brambilla, A.~Pineda, J.~Soto, A.~Vairo, {The QCD potential at $O(1/m)$},
  Phys.Rev. D63 (2001) 014023.
\newblock \href {http://arxiv.org/abs/hep-ph/0002250}
  {\path{arXiv:hep-ph/0002250}}.

\bibitem{Pineda:2000sz}
A.~Pineda, A.~Vairo, {The QCD potential at $O(1/m^2)$: Complete spin dependent
  and spin independent result}, Phys.Rev. D63 (2001) 054007.
\newblock \href {http://arxiv.org/abs/hep-ph/0009145}
  {\path{arXiv:hep-ph/0009145}}.

\bibitem{Jugeau:2003df}
F.~Jugeau, H.~Sazdjian, {Bound state equation in the Wilson loop approach with
  minimal surfaces}, Nucl.Phys. B670 (2003) 221.
\newblock \href {http://arxiv.org/abs/hep-ph/0305021}
  {\path{arXiv:hep-ph/0305021}}.

\bibitem{Sazdjian:2007ng}
H.~Sazdjian, {Integral equation for gauge invariant quark two-point Green's
  function in QCD}, Phys.Rev. D77 (2008) 045028.
\newblock \href {http://arxiv.org/abs/0709.0161} {\path{arXiv:0709.0161}}.

\bibitem{Sazdjian:2010ku}
H.~Sazdjian, {Spectral properties of the gauge invariant quark Green's function
  in two-dimensional QCD}, Phys.Rev. D81 (2010) 114008.
\newblock \href {http://arxiv.org/abs/1003.5099} {\path{arXiv:1003.5099}}.

\bibitem{Dyson:1949ha}
F.~Dyson, {The S matrix in quantum electrodynamics}, Phys.Rev. 75 (1949) 1736.

\bibitem{Schwinger:1951ex}
J.~S. Schwinger, {On the Green's functions of quantized fields. 1.},
  Proc.Nat.Acad.Sci. 37 (1951) 452.

\bibitem{Schwinger:1951hq}
J.~S. Schwinger, {On the Green's functions of quantized fields. 2.},
  Proc.Nat.Acad.Sci. 37 (1951) 455.

\bibitem{Alkofer:2000wg}
R.~Alkofer, L.~von Smekal, {The infrared behavior of QCD Green's functions:
  Confinement, dynamical symmetry breaking, and hadrons as relativistic bound
  states}, Phys.Rept. 353 (2001) 281.
\newblock \href {http://arxiv.org/abs/hep-ph/0007355}
  {\path{arXiv:hep-ph/0007355}}.

\bibitem{Fischer:2006ub}
C.~S. Fischer, {Infrared properties of QCD from Dyson-Schwinger equations},
  J.Phys. G32 (2006) R253.
\newblock \href {http://arxiv.org/abs/hep-ph/0605173}
  {\path{arXiv:hep-ph/0605173}}.

\bibitem{Corrigan:1978zg}
E.~Corrigan, B.~Hasslacher, {A functional equation for exponential loop
  integrals in gauge theories}, Phys.Lett. B81 (1979) 181.

\bibitem{Durand:1979sw}
L.~Durand, E.~Mendel, {Functional equations for path dependent phase factors in
  Yang-Mills theories}, Phys.Lett. B85 (1979) 241.

\bibitem{'tHooft:1973jz}
G.~'t~Hooft, {A planar diagram theory for strong interactions}, Nucl.Phys. B72
  (1974) 461.

\bibitem{'tHooft:1974hx}
G.~'t~Hooft, {A two-dimensional model for mesons}, Nucl.Phys. B75 (1974) 461.

\bibitem{Kazakov:1980zi}
V.~Kazakov, I.~Kostov, {Nonlinear strings in two-dimensional U($\infty$) gauge
  theory}, Nucl.Phys. B176 (1980) 199.

\bibitem{Kazakov:1980zj}
V.~Kazakov, {Wilson loop average for an arbitrary contour in two-dimensional
  U(N) gauge theory}, Nucl.Phys. B179 (1981) 283.

\bibitem{Bralic:1980ra}
N.~E. Bralic, {Exact computation of loop averages in two-dimensional Yang-Mills
  theory}, Phys.Rev. D22 (1980) 3090.

\bibitem{Wightman:1956zz}
A.~Wightman, {Quantum field theory in terms of vacuum expectation values},
  Phys.Rev. 101 (1956) 860.

\bibitem{Schweber:1961zz}
S.~S. Schweber, {An introduction to relativistic quantum field theory}, Row,
  Peterson and Co., Evanston, 1961.

\bibitem{'tHooft:1973pz}
G.~'t~Hooft, M.~Veltman, {Diagrammar}, NATO Adv.Study Inst.Ser.B Phys. 4 (1974)
  177.

\bibitem{Politzer:1976tv}
H.~D. Politzer, {Effective quark masses in the chiral limit}, Nucl.Phys. B117
  (1976) 397.

\bibitem{Kallen:1952zz}
G.~Kallen, {On the definition of the renormalization constants in quantum
  electrodynamics}, Helv.Phys.Acta 25 (1952) 417.

\bibitem{Lehmann:1954xi}
H.~Lehmann, {On the properties of propagation functions and renormalization
  constants of quantized fields}, Nuovo Cim. 11 (1954) 342.

\end{thebibliography}






% \end{thebibliography}

\end{document}